# IMPROVING DATA USE AND PARTICIPATORY ACTION AND DESIGN TO SUPPORT DATA USE: THE CASE OF DHIS2 IN RWANDA

Scott Russpatrick, University of Oslo, scottmr@ifi.uio.no

Magnus Li, University of Oslo, magl@ifi.uio.no

Jørn Braa, University of Oslo, jornbraa@gmail.com

Alexander Bruland, University of Oslo, alexbrul@ifi.uio.no

Mikael Olsen Rodvelt, University of Oslo, mikaelor@ifi.uio.no

Andrew Muhire, HISP Rwanda, muhireandrew@gmail.com

Adolphe Kamungunga, HISP Rwanda, akamungunga@hisprwanda.org

Kai Vandivier, University of Oslo, kai@dhis2.org

Silvia Masiero, University of Oslo, silvima@ifi.uio.no

Peichun Wu, University of Oslo, peichunw@ifi.uio.no

Peter Biro, University of Oslo, peterbir@ifi.uio.no

Stian Rustad, University of Oslo, stiaru@ifi.uio.no

**Abstract.** This article reports from an ongoing 'evaluation for improvement' action research and participatory design project in Rwanda, where the aim is to improve data use practices and the capabilities of the District Health Information Software 2 (DHIS2), an open source health information management platform, to support data use. The study of data use at health facility and district level showed that while data was used routinely at, for example, monthly coordination meetings, the DHIS2 dashboards and other analytical tools were in limited use because users preferred to use Microsoft Excel for data analysis and use. Given such findings, a major focus of the project has been directed towards identifying shortcomings in data use practices and in the software platform and to suggest, design and eventually implement changes. While the practical work on implementing improvements have been slow due to the COVID-19 pandemic, the suggested design improvements involve many levels of system design and participation, from the global core DHIS2 software team, the country DHIS2 team and local app development, the Rwanda Ministry of Health, and health workers at local level.

## 1. INTRODUCTION

This article is about an ongoing participatory design (PD) project in Rwanda, which has the twofold aim to both assess and improve the use of data and the use and usefulness of the District Health Information Software 2 (DHIS2) software platform (https://dhis2.org/), which is the standard software tool in the health services in the country. A team of core DHIS2 developers and students from the Design lab at the University of Oslo are working with the team from the Rwanda Ministry of Health in responding to users requirements by developing apps and DHIS2 features in direct interaction with DHIS2 users in the Rwanda health services. Given this background, this article will describe and discuss some key challenges related to multi-levelled and local / global participatory design using the DHIS2 open source software platform. While DHIS2 has a flexible metadata structure and a certain plasticity allowing for local PD and customization, many local requirements cannot be accommodated by the generic core DHIS2 platform. The way to address user requirements which cannot be met by the core platform, as described in [1], has been to add needed new features to the roadmap for core DHIS2 development. As it is demonstrated by the case of Rwanda, core development is too slow for practical PD and end users requests, and all user requirements do not necessarily fit in as new features and apps in the platform.

Another aim with this article is to develop approaches whereby PD can be better applied at large global scale addressing what Bødker and Kyng [2] label 'big issues'. They argue that PD is now too much confined to small and insignificant pilot projects with a focus on direct collaboration between users and designers in co-design processes to engage with usability and human-computer interaction (HCI), leading to a focus on 'small issues', in contrast to 'big and important ones'. In their aim to revitalize and revise PD to better address and influence big issues, they use HISP [3] as their inspiration to revitalize PD for a new type of PD (ibid.). While HISP's aim to improve data use for better health in the global south is an obviously 'big issue', important inspirations, according to Bødker and Kyng, are HISP's self sustaining global network, ongoing participatory processes in support of local/local and local/central cooperation on data use and thereby facilitating scaling up (ibid.).

Improvement in the DHIS2 technical design is going on both at local and global levels. While the DHIS2 country team is configuring the DHIS2 to suit the identified requirements within the generic capabilities of the platform, other requirements need involvement by the DHIS2 global team. Using DHIS2 as a platform and developing apps to address specific features and user requirements represent a third 'in between' opportunity. Apps can be, and are, developed both locally in Rwanda as well as in other countries and shared. We present examples of all these types and levels of responding to user requirements and discuss the findings related to challenges and opportunities in engaging - and being dependent upon - both local and global system developers in participatory design going on in a real world context where not all factors are controlled, such as for example the ongoing pandemic.

This research will contribute to the theory and practices of participatory design and action research in a global-local multi-level perspective [4,5,6], and to the role of app development to provide more agile responses to identified user needs [7].

## 2. BACKGROUND DHIS2

Since 2012, the Ministry of Health has been implementing the DHIS2 powered Health Management Information System (HMIS) to manage countrywide health related data. The HMIS system includes health facility level data from all health service areas and programs. Although much effort has been invested in improving the quality of data, timeliness of routine reporting as well as data use for informed decisions, supervisions and assessments conducted have shown that data use at health facility and district levels is far from optimal and that the various analytic tools in the DHIS2 were in limited use. These gaps and challenges faced had to be addressed and documented to guide the next steps.

As DHIS2 is designed and developed to be used across multiple countries and user organizations, design unfolds on multiple levels. A core team is responsible for the design and development of the generic software features and caters for the wide audience of use-cases, forming a level of generic design. Further, for each implementation, the DHIS2 is configured and customized according to the more specific circumstances, and this forms a process of implementation-level design [8]. DHIS2 has a flexible metadata structure and can be configured for a variety of needs and system requirements [4] When addressing user needs, new functionalities and requirements not covered by the generic features can basically be met in three ways; 1) customisation of the generic functionalities in DHIS2, done locally, 2) request changes in the core platform through interacting with global teams, and 3) develop specific apps, which can be developed locally, regionally and globally.

The objectives of the research reported in this paper are to evaluate and document gaps and challenges faced by local level users at districts and health facilities, to use DHIS2 analytics apps for data analysis and other features in their use of data for management, planning and service delivery. Furthermore, having identified the requirements, interventions will be designed and implemented on the basis of the three approaches of local customisation, DHIS2 core development and /or app development.

Given the above objectives, a team from the University of Oslo in collaboration with HISP Rwanda under the guidance of the MoH/HMIS conducted an assessment on the status of data use in Rwanda. We used a data use checklist developed by the University of Oslo to guide countries in assessing the level of use of data and dashboards.

## 3. DHIS2 AND DIGITAL PLATFORMS

Tiwana defines a software platform as a "software-based product or service that serves as a foundation on which outside parties can build complementary products or services" [9]. Tiwana also argues that platforms must be multisided, meaning they bring together two or more actors or groups of platform users, such as end users and app developers, similar to our case of DHIS2 and app development. Henfridsson and Bygstad argue that in order to analyse digital platform developments, the focus need to be on the boundary resources of the platform and not on the platform itself [10]. In the literature are two types of boundary resources, technology and knowledge [11, 12]. Technology boundary resources, provided by the platform owner, enable 3rd party developers to create new applications and features through technical resources such as APIs, SDKs, code libraries, UX templates and standards, etc [13, 14, 15]. Knowledge boundary resources provide practical knowledge necessary for 3rd party developers to access and utilize the technical boundary resources. There are guidelines, programming tutorials, information portals, courses, development roadmaps, workshops, and co-innovation projects [15, 16, 17]

In our case, these resources are made up of the technical: the DHIS2 app development environment, tools for mock-up development, and resource libraries, and the knowledge: zoom and shared google doc environment facilitating the interaction and collaboration between app developers in Oslo and users and developers in Rwanda. To further emphasise such socio-technical understanding of platforms we use Gawer's approach to bridge two theoretical perspectives: economics, which see platforms as double sided markets that have both demand and supply-side users and with network effects between and within them, and engineering, which sees platforms as technological architecture. This leads Gawer to a new conceptualization of platforms as evolving organisations or meta organisations [18], which is in line with how we conceptualise our case of global participatory design and app development within and around the DHIS2 platform.

DHIS2 is an example of an innovation *platform* that provides the foundation for which applications or components can be built upon. A classic example is the Android operating system, where a stable core is maintained that allows for periphery application to be supported. Mandel [19] describes this as application economies, or "a collection of interlocking innovative ecosystems where each ecosystem consists of a core ecosystem, which creates and maintains a platform and an app marketplace." In the application economy multiple individuals, groups and organizations – eg. developers, companies, or governments- are able to create, launch, and maintain their own applications. Innovation platforms enable a large number of innovators to develop complementary applications or services within the platform ecosystem by providing technical foundational elements [20]. Applications can be highly specific to a single end-user, such as an application that aids community health workers in Zambia in diagnosing Malaria, or highly generic to a broad range of end-users, such as Whatsapp [21].

DHIS2 has a core database and API developed and maintained by the Health Information System Programme (HISP) headquartered at the University of Oslo. The core development team also develops and maintains a suit of "core" generic applications that are intended to be the minimal tools necessary for an HMIS. These are: data capture applications, analytics applications, such as dashboards, pivot tables, charts, and maps. There are also data quality application as well as meta-data configuration and user management applications. These applications reuse common components and exist on top of a stable application programming interface (API) forming a layered, modular architecture. Beyond the core is a proliferation of locally developed applications that are developed with little or no involvement from the core development team. These periphery applications can be generic and, thus, more reusable across countries and contexts or highly specialized for a specific end-user or function [7].

## 4. METHODS

The research is carried out in two phases. While the first 'pre pandemic' phase was based on fieldwork in Rwanda the second phase was carried out during the pandemic and consisted of online meetings and collaborative work between users in the health services and technical HISP team in Rwanda and core DHIS2 developers and students in the Design lab at the University of Oslo.

**First phase -2019**
The research approach has included both qualitative fieldwork and user interaction at district and health facility levels using a semi-structured interview guide and a quantitative approach using questionnaires. A WhatsApp group has been created bringing together the Data Managers of Health centers and District Hospitals to freely and openly keep sharing their experiences and providing feedback to inform improvements in DHIS2. Requirements identified through the field visits, questionnaires and the WhatsApp group were both addressing needs for global level changes in DHIS2, i.e. input to new releases, and requests to the local DHIS2 team.

The team visited three district offices and 6 health facilities in Kigali, the capital, and in Gisenyi, in the north-west of the country, during three weeks September 2019. Here focus group discussions were conducted and digital information practices were investigated hands-on. A series of planning meetings organized at National level with the global DHIS2 team (HISP Rwanda, Uio, MOH) to ensure more issues are identified and feedback are shared. The team also attended monthly meetings (DHMTs, coordination meetings, quality assurance meetings) and in-depth interviews with data managers and facility managers to identify best practices, challenges, and user requests. We conducted group discussions with key players at health facilities and district offices following a check-list which included investigation and discussion of the computerised tool they used, such as reports, dashboards, graphs and tables. The questionnaire was shared with additional 36 District Hospital Data Managers who were asked to provide feedback and request new features to support their data use needs.

**Phase 2 - November 2020-March 2021**
The overall cyclic research approach was initially planned is as follows:
1) Identify situations and venues for routine data use and use of DHIS2 at facility and district levels and assess data use with a particular focus on assessing the way the DHIS2 software platform and other digital tools are used, or not used, to support and enhance data use,
2) Identify shortcomings in how DHIS2 is used and or designed to support data use and users and suggest improvements in software design and use practices,
3) Implement and evaluate suggested changes, and repeat the cycle of through 1, 2 and 3. This last cycle of implementing changes, however, has been somewhat halted due to the COVID-19 pandemic.

The research was halted during about 10 months in 2020 due to the COVID-19 pandemic. It was eventually restarted as an online collaboration using zoom and teams tools for communication and also a Mock-up tool facilitating participation and collaboration on specification of the app to be developed. Students and staff from the Design lab at uio, as well as staff from the global DHIS2 team made up the UiO part of the collaboration while the Rwanda team consisted of MoH and HISP Rwanda staff and information system officers from selected District hospitals. These district officers are responsible for information management and use in the district which is typically made up of about 20 health centres and one district hospital. Support and improvement of data management and use at district level and in hospitals and health centres are key targets for the project. During February - March, 6 online meetings between Oslo and Rwanda were conducted, with about 15-20 participants, and where use cases and requirements were presented and solutions discussed. A Mock-up tool was used as a means to communicate potential solutions between the developers in Oslo and users in Rwanda. A team from HISP Tanzania did also participate in the meetings. They are experienced app developers and the intention is to include them in the practical app development.

To collaboratively develop potential prototypes/mockups, the UiO students suggested a mockup tool called Figma. Figma is supported both in browsers and as a desktop application, and allows for real-time synchronization across different instances. Such a feature serves as a key element in supporting the development process in a cross-country, cross-institutional setting, when also being restricted by online-tools as the sole means for communication. The approach taken in this project when developing concepts of a minimal viable product (MVP), is as an attempt to incorporate a more agile tendency in the development process of current and future software for DHIS2. During this phase, the HISP teams and representatives from UiO have iteratively met to discuss the potential of the current prototype, and provided feedback to guide the improvement in the next iteration. As a result, the development process has engaged the collaboration in a more agile fashion, benefiting both the developers and stakeholders.

## 5. CASE: ONLINE GLOBAL PARTICIPATORY DESIGN

We made contact through Google Meet between the teams from Rwanda, Tanzania and UIO. Rwanda and Tanzania described their problem and the need for a Data Form App. The need was clear, but the user stories caused confusion and questions. We discussed two cases as possible solutions;

1. Create a feature within the existing Data Set Report App, which allows to combine different data sets from Provincial Hospital report forms, into customised formats.
2. Create a DHIS2 App. (Selected approach)
    1. Create Report format (Admin/ Super User ) in accordance to MOH-HMIS agreed formats.
    2. Share Report format to respective groups of users for the responsible teams to generate the report with respective datasets that correspond to the particular level.
    3. Generate the consolidated report once the data manager confirms all the data sets have been reported in.
    4. Reports can then be downloaded as xls or pdf.

It was discussed that the use case was similar to an existing Dashboard feature. The dashboards have printing options and are made for decision making. The teams agreed to proceed on the second approach. Both Rwanda and Tanzania showed great interest in the project and understood the need for generic development and collaboration. Further weekly meetings were used to summarise and comment, while tasks were spread between teams. The first meeting discussing the use case had 27 invited members, representing Health Ministry of Tanzania, HISP Rwanda, HISP Tanzania, UIO team and students.

The weekly meetings had around 18 attendees. They have been used to present work and obtain feedback from the whole team. These meetings are very important, as they provide feedback from the DHIS2 team. The will and understanding of a generic development exists for all the participants, but it's lacking knowledge and experience of the DHIS2 platform resources. Feedback has been very important, to help the team create a plan that involves the correct use of DHIS2 resources and design. For example, using the DHIS2 components already in the prototype ensures generic development further in the project. As it can seem obvious once it's brought up, it was not a general understanding before feedback was given.

There have been four meetings between HISP Rwanda team and UIO students, with 7 to 20 participants. First meeting had some confusion due to time zones, but the following meetings got more attendees from Rwanda and Tanzania. HISP Rwanda made a Data Flow Diagram, and these meetings are currently used to create an interactive prototype and to create a detailed roadmap.

It has been a total of four meetings. One discussing the use case, one discussing the proposed flow diagram from HISP Rwanda, one between UIO students, Rwanda and Tanzania, and another weekly meeting presenting roadmap and prototype drafts. The process is moving fast with a lot of involvement from all sides.

## 6. FINDINGS

An important finding is that all districts in Rwanda are conducting monthly meetings where health facility and district managers and data managers come together at the district level in a meeting called 'coordination meeting' and evaluate last month's data, look at trends and discuss needed action. This routine event is identified as a key area for DHIS2 to provide information support. However, while the finding was that the level of use of data from the DHIS2, for example at the coordination meetings, was maybe surprisingly high, use of DHIS2 for analysing and presenting data, for example through dashboards, was surprisingly poor. The data managers preferred to download the data from DHIS2 to Excel and analyse and present the data using Excel as the preferred tool. Given this finding, ways to improve the DHIS2 as a tool for data managers and other users and to improve data practices using the DHIS2 became the focus for the participatory design and action research part of the project.

One example of why users preferred Excel to DHIS2 is from a posting on the WhatsApp group where a data manager explained that he needed to be able to add comments as text and colours as, for example, red for poor performance, in a table of an overview of health facilities' performance, as in the figure.

**Fig. 1.** Excerpt – health facility data presentation, Rwanda

Adding text and colours in a table in a web based system such as DHIS2 is not straightforward. This is an example of an user request that requires involvement of the core DHIS2 development team. Other identified requirements are also addressing core DHIS2 features and global involvement, for example to be able to build reports combining visuals and text and which is populated with data from DHIS2.

The results from the questionnaires confirmed the findings from the field visits, with four notable pointers:

- Most of the Health Facilities and district offices download data from DHIS2 pivot table to Excel for doing their reports by-passing DHIS2 limitations like to be able to manipulate charts and tables, to add text and colours etc.
- Nearly all know how to make pivot tables, charts, maps, and dashboards in DHIS2, but due to the limitations in DHIS2, pivot tables are generated as a means to download data to Excel.
- An additional reason for downloading to Excel was that target population denominator data were not available in DHIS2 for sub-units like health posts under the health centres. The local users will know the figures and then it is easier to enter them directly in Excel.
- Most to all respondents wanted the ability to: mix colours and text in tables as in the figure above; share dashboards to predefined groups of users; print dashboards and use as reports; build reports where visuals are dynamically generated and text can be added e.g. for monthly reports; generate pdf of monthly reports to print out and sign for archiving

## 7. ADDRESSING ISSUES THROUGH DESIGN: GENERIC AND IMPLEMENTATION-LEVEL DESIGN AND DEVELOPMENT

The evaluation revealed several important requirements for new features and performance that would need to be addressed before the DHIS2 platform may effectively support the identified data use situation. Some of the requested improvements can be carried out locally on the level of implementation through customisation of the DHIS2, such as the ability to print out reporting forms dynamically filled with data for signature and archiving. The local HISP Rwanda team is capable of optimally configuring and customising the DHIS2 platform, but many requirements go beyond the current customization capabilities of the DHIS2. These can either be addressed through design on the generic level by the global developer team, or on the level of implementation by building custom apps.

**Addressing issues on the generic level of design and development**

DHIS2 and dashboard enhancements requested need to be considered for in the core platform development. This includes among the other features that can allow/support print and downloading of dashboard or dashboard items. A community participatory design approach ) is employed for this in which the requirements are gathered from the field via a lead implementer in HISP Rwanda. Those requirements are then passed to the platform owner through which collates all field requests. Representatives from all HISPs then vote on the collated requirements to produce a development roadmap that is then produced by the core developers. The DHIS2 platform owners must take into account what is technically achievable for the core developers to accomplish. In this case study from the direct field requirements coming from Rwanda the core developers appreciated that dashboard download and printing was a realistically achievable for the next DHIS2 core release; however, editing cell color and adding comments in a pivot table required an extensive expansion of the current pivot tables functionality and would not be possible in the release. It also is not possible for the core development team with limited resources and many demands to reproduce Excel into DHIS2, so the priority of this request is relatively low given its technical achievability.

Given the inherent inertia and understandable slow pace of change of the DHIS2 platform, local participatory design efforts need to look elsewhere to be able to respond fast enough to users requests.

**Implementation-level participatory design**

Implementation-level participatory design processes require a constant agile response by way of new features and solutions to be fed back and tested by the users. When, as in our case, the generic features in DHIS2 cannot fully be used to respond to user requirements and the generic-level design process of the global developer team is too slow to achieve, app development is a way to bridge the gap. Rwanda has already developed several apps and have both expertise and experience in this regard, for example:

- When implementing the DHIS2 Tracker for the immunisation program vaccine stock management was not part of the package, and was immediately identified as a gap. HISP Rwanda then developed an app for managing vaccine stocks and requests for vaccines and distribution from available sources; central and regional warehouses and hospitals. This app is now being developed as a generic central DHIS2 app.
- An app including a certificate for having tested negative for COVID-19 and linked to laboratories was developed in collaboration with HISP Uganda and is now used by truck drivers and other travelers crossing the borders between the two countries. Tanzania, neighbouring both countries, is now adopting and adapting this app.

Analysing the requirements and responses from the DHIS2 core team, we see that they fall into two levels:
1) Generic level design and apps by DHIS2 core team apps. Very few user requirements identified during the process can be met by generic features and apps within a participatory timeframe, as fast responses are required. Ability to print dashboards is one example of a generic app which will be made available soon.
2) Implementation level design, as described in Rwanda. Three options are available; 2.1) customise solutions based on generic features, and 2.2) request generic features from the core team, which we have seen may be too slow. 2.3) Building custom apps. As with requesting features from the core team, the other options also imply challenges. Customizing generic apps requires an understanding of their existing codebase and will later result in issues with updating to new versions released by the core team. Building custom apps provide a flexible alternative, yet development and future maintenance must be done locally, which could be costly.

Of the requirements identified, the ability to print reporting forms filled dynamically with data is a 'low hanging' fruit. This is a most wanted feature and can be done by customisation or a small app. Other requirements, such as a more advanced report builder than the previous example need to be addressed through building custom apps. Rwanda and regional partners have proven their ability to build apps, but as described, it will involve development and maintenance work locally.

**Boundary Resources: Building local and global knowledge together**

Boundary resources, both knowledge and technological, have been defined in the existing literature to be produced by the platform owner, University of Oslo, for the 3rd party application developer, HISP Rwanda. However, in this case study we see that knowledge boundary resources through close, interactive interactions are developed simultaneously and complementary on by both the platform owner and the 3rd party developer.

Specifically, by HISP Rwanda communicating needs and user-stories, the UiO product management team was able to define new features that could be added to the core DHIS2 feature sets. However, it also illuminated the necessity for features that the core UiO development team would not develop, for example reproducing many excel features in DHIS2. This understanding of what UiO would and would not develop gave HISP Rwanda the knowledge needed to build a new application locally that would satisfy those unmet requirements. For UiO, knowing the user stories added the product management team to formulate a more explicit roadmap that is communicated to all DHIS2 implementers globally. The more clearly defined roadmap then gives other 3rd party developers the information they need to know if they will have to make a new application or feature based on their unique needs that UiO is not planning to develop.

This case-study shows that generation of knowledge boundary resources is not solely top-down, but can be bidirectional, informed from direct engagement between platform owner and 3rd party developer.

## 8. CONCLUSION - AND FURTHER RESEARCH

The traditional means of conducting participatory design using the DHIS2 platform has been to address needs through generic features, and when that is not possible, to request changes by the core team. A conclusion of this research, however, is that participatory design and addressing emerging needs in a mature DHIS2 user country such as Rwanda, where the generic features are already exhausted, needs to rely on custom app development. In a global perspective, then, local custom apps will inform development of core generic apps and features. There are however challenges related to local apps development, such as for example, maintenance. Our further research will explore challenges and opportunities for a custom app development strategy.

The contributions of the research reported in this paper are in 1) developing practical action research approaches synthesising research and practice in developing information systems in developing country settings [8, 5], and 2) providing approaches and analysis of multi-levelled participatory design and interaction across local site, country and global levels [4, 7, 10, 6]. At the practical level contributions are on how tension can be eased between slow global responses on generic platform features and local needs through local app development.